\documentclass[a4paper]{article}
\usepackage{fullpage,color}

\usepackage{amsmath,amsbsy,amssymb,cite}

\title{On a direct approach to quasideterminant solutions of a noncommutative modified KP equation}
\author{ C. R. Gilson, J. J. C. Nimmo and C. M. Sooman \\
Department of Mathematics, \\
University of Glasgow \\
Glasgow G12 8QW, UK}
\date{}

\newtheorem{thm}{Theorem}[section]

\newtheorem{defn}{Definition}[section]

\numberwithin{equation}{section}

\begin{document}
	
\maketitle

\begin{abstract}
A noncommutative version of the modified KP equation and a family of its solutions expressed as
quasideterminants are discussed.  The origin of these solutions is explained by means of Darboux transformations and the solutions are verified directly.  We also verify directly an explicit connection between quasideterminant solutions of the noncommutative mKP equation and the noncommutative KP equation arising from the Miura transformation.
\end{abstract}

\section{Introduction}

Recently, there has been much interest in several noncommutative integrable systems \cite{MR2033217,MR2017171,Wang:Wadati,Dimakis:Muller-Hoissen2006,MR2114103,MR2147176}.  In the paper by Etingof \textit{et al} \cite{gelfand-retakh-paper-with-appendix-on-this-topic}, it was shown that solutions of the noncommutative KP equation (ncKP) could be expressed as quasideterminants.  Quasideterminant solutions of noncommutative integrable systems can often be obtained from Darboux transformations \cite{quasideterminants:main, MR1634885}.  These concepts are elaborated on in \cite{claire:jon} where two families of solutions of the ncKP equation were presented which were termed quasiwronskians and quasigrammians. The origin of these solutions were explained by Darboux and binary Darboux transformations. The quasideterminant solutions were then verified directly using formulae for derivatives of quasideterminants (see also \cite{MR2321659}). In this approach, the nature of the noncommutativity is not specified so that the results presented were valid for, for example, the noncommutative Moyal star product and the matrix or quaternion versions of the KP equation.

In the present paper, we follow this concept of noncommutativity to find quasideterminant solutions of a
noncommutative version of the modified KP equation (ncmKP).  The ncmKP hierarchy \cite{kupershmidt-book} can be constructed in the spirit of Gelfand-Dickii theory \cite{Gelfand:Dickii} and the ncmKP equation extracted from it using a change of variables given in \cite{Wang:Wadati}. The origin of the solutions of ncmKP are explained by means of Darboux transformations of the pseudo-differential operator used to construct the hierarchy.  Obtaining Darboux transformations in this manner is reminiscent of the approach in \cite{Oevel:Rogers} where the solutions are expressed as ratios of wronskian determinants.  We extend the concepts in \cite{Oevel:Rogers} to the ncmKP hierarchy and find a class of quasiwronskian solutions obtained by iteration of the gauge-transformed pseudo-differential operator for the ncmKP hierarchy, interpreting this process as a Darboux transformation.  It is then shown that these solutions can be verified directly using formulae for derivatives of quasideterminants and some related identities.  In \cite{Oevel:Rogers} it is shown that a Miura transformation between the (commutative) KP and mKP equations can be obtained from a gauge transformation of the pseudo-differential operator used to construct the (commutative) KP hierarchy by comparing this operator to the pseudo-differential operator used in the construction of the (commutative) mKP hierarchy. In the book by Kupershmidt \cite{kupershmidt-book}, a noncommutative Miura transformation between the ncKP and ncmKP equations was given.  Here we present a noncommutative Miura transformation analogous to that given in \cite{Dimakis:Muller-Hoissen2006}.  We also give the explicit connection between the quasideterminant solutions of ncKP and ncmKP that is described by the Miura transformation.

The present work requires the use of some elementary properties of quasideterminants, which we shall recall in Section \ref{sec3}. For a complete treatment of quasideterminants, the reader should refer to the original papers \cite{Gelfand1991, MR1613523, quasideterminants:main}.

\section{Noncommutative mKP hierarchy}\label{sec2}
In this section, we construct the ncmKP hierarchy in the spirit of Gelfand-Dickii theory~\cite{Gelfand:Dickii, gelfand-retakh-paper-with-appendix-on-this-topic}.  A pseudo-differential operator $L$ is defined by
\begin{align*}
	L &= \partial_x + w + w_1 \partial^{-1}_x + w_2\partial^{-2}_x + w_3 \partial^{-3}_x + \cdots,
\end{align*}
where $w$ and $w_s (s=1,2,\ldots)$ do not necessarily commute and depend on $x$ and $t_q (q=1,2, \ldots)$, and $\partial_x^i$ denotes the $n$th partial derivative operator $\frac{\partial^i}{\partial x^i}$.  As in standard, commutative Sato theory, we define the ncmKP hierarchy as
\begin{align}
L_{t_q} &= [P_{\geq 1} (L^q), L], \qquad q = 1,2, \ldots,  \label{commutater1}
\end{align}
where
\begin{eqnarray*}
  P_{\geq 1} \left( \sum_i w_i \partial_x^i \right) = \sum_{i \geq 1} w_i
  \partial_x^i,
\end{eqnarray*}
denotes projections of powers of the operator $L$ onto the differential part.  The first three such projections are
\begin{align*}
    P_{\geq 1}(L) &= \partial_x, \\
    P_{\geq 1}(L^2) &= \partial^2_x + 2w \partial_x, \\
    P_{\geq 1}(L^3) &= \partial^3_x + 3w \partial^2_x + 3(w_x + w^2 + w_1) \partial
    x.
\end{align*}
Thus, via the evolution equation (\ref{commutater1}), we obtain the ncmKP hierarchy:
\begin{align}
   L_{t_1} &= [P_{\geq 1}(L),L] \Leftrightarrow \left\{ \begin{array}{rll} w_{t_1} & =  & w_x, \\ w_{1t_1} &
    = & w_{1x}, \\ w_{2t_1} & = & w_{2x}, \\ {} & \cdots & {} ,\end{array} \right. \label{hierarchy7}\\
    L_{t_2} &= [P_{\geq 1}(L^2),L] \Leftrightarrow \left\{ \begin{array}{rll} w_y & = & w_{xx} + 2w_{1x}
    + 2ww_x + 2[w,w_1], \\
    w_{1y} & = & w_{1xx} + 2w_{2x} + 2w_1 w_x + 2ww_{1x} + 2[w,w_2], \\
    w_{2y} & = & w_{2xx} + 2w_{3x} + 2ww_{2x} + 4w_2w_x - 2 w_1 w_{xx} + 2[w,w_3], \\
    w_{3y} & = & w_{3xx} + 2w_{4x} + 2ww_{3x} + 6w_3w_x - 2w_1w_{xxx} - 6w_2w_{xx} \\ {} & {} & + \, 2[w,w_4], \\
    {} & \cdots & {} ,\end{array} \right. \label{hierarchy8} \\
    L_{t_3} &= [P_{\geq 1}(L^3),L] \Leftrightarrow \left\{ \begin{array}{rll} w_t & =  & w_{xxx} + 3w_{1xx} + 3w_{2x} + 6ww_{1x}
    + 3w_1w_x + 3w_xw_1 \\ {} & {} & + \, 3ww_{xx}  + 3 w_x^2 + 3w^2w_x + 3[w^2,w_1] + 3[w,w_2], \\
    {} & \cdots & {} ,\end{array} \right. \label{hierarchy9}
\end{align}
where we have set $t_2 =y$ and $t_3 = t$.  The term $2[w,w_1]$ in the first component of (\ref{hierarchy8}) prevents us from recursively expressing the fields $w_s (s=1,2, \ldots)$ in terms of $w$ and its $x$- and $t_q$-derivatives. However, using the second component of (\ref{hierarchy8}) and the first component of (\ref{hierarchy9}), we obtain
\begin{align}
    2w_t - 2w_{xxx} - 3w_{1xx} - 6ww_{1x} - 3w_{1y} - 6w_xw_1 - 6ww_{xx} - 6w_x^2 - 6w^2w_x - 6[w^2, w_1] = 0.
    \label{ncmKPbare}
\end{align}
To eliminate the field $w_1$, we make the change of variables $w_1 = - \frac{1}{2}(w_x + w^2 - W)$.  Thus, from
the first component of (\ref{hierarchy8}), and from (\ref{ncmKPbare}), we obtain the following equations:

\begin{align}
    -4w_t + w_{xxx} - 6ww_xw + 3W_y + 3[w_x,W]_+ - 3[w_{xx},w] -
    3[W,w^2] = 0, \label{NCmKPequation} \\
    W_x - w_y + [w,W] = 0. \label{ncmKpextra}
\end{align}
Equations (\ref{NCmKPequation}, \ref{ncmKpextra}) form the ncmKP equation, in a slightly different but equivalent form to that studied in \cite{Wang:Wadati}.  This is found from a different perspective in \cite{Dimakis:Muller-Hoissen2006}.  Equation (\ref{ncmKpextra}) can be satisfied identically by introducing the change of variables $w = - f_x f^{-1}$, and $W = - f_yf^{-1}$ (see also \cite{Wang:Wadati}) where $f=f(x,t_q)$ is invertible but is not assumed that $f$ and its derivatives commute.

\section{Quasideterminants}\label{sec3}

Quasideterminants were introduced by Gelfand \textit{et al} in the early 1990s \cite{Gelfand1991}.  Here we give the basic definitions and a summary of the results from this theory that we will use.  An $n\times n$ matrix $A$ over a not necessarily commutative ring $\mathcal{R}$ has, in general, $n^2$
quasideterminants.  We denote each quasideterminant by $|A|_{ij}, 1 \leq i,j \leq n$. Let $A^{ij}$, which we assume is invertible, denote the matrix obtained from $A$ by deleting the $i$th row and $j$th column.  Let $r_k^j$ be the row vector obtained from the $k$th row of $A$ by deleting the $j$th entry and let $s_l^i$ be the column vector obtained from the $l$th column of $A$ by deleting the $i$th entry. Then $|A|_{ij}$ exists and
\begin{align}
    |A|_{ij} &= a_{ij} - r_i^j(A^{ij})^{-1}s_j^i.  \label{quasidet1}
\end{align}

We shall henceforth adopt an alternative notation for quasideterminants by boxing the leading element $a_{ij}$. More generally, for a block matrix, we can define

\begin{align*}
    \left| \begin{array}{cc} A & B \\ C & \fbox{$d$} \end{array} \right| &= d-CA^{-1}B,
\end{align*}
where $d \in \mathcal{R}$, $A$ is a square matrix over $\mathcal{R}$ of arbitrary size and $B,C$ are column and
row vectors over $\mathcal{R}$ of compatible lengths.

\subsection{Homological relations}

It is shown in \cite{quasideterminants:main} that quasideterminant row and column homological relations can be written as

\begin{align}
    \left| \begin{array}{ccc} A & B & C \\ D & f & g \\ E & \fbox{$h$} & i \end{array} \right|
    &= \left| \begin{array}{ccc} A & B & C \\ D & f & g \\ E & h & \fbox{$i$} \end{array} \right|
        \left| \begin{array}{ccc} A & B & C \\ D & f & g \\ 0 & \fbox{$0$} & 1 \end{array} \right|
\end{align}
and
\begin{align}
    \left| \begin{array}{ccc} A & B & C \\ D & f & \fbox{$g$} \\ E & h & i \end{array} \right|
    &= \left| \begin{array}{ccc} A & B & 0 \\ D & f & \fbox{$0$} \\ E & h & 1 \end{array} \right|
        \left| \begin{array}{ccc} A & B & C \\ D & f & g \\ E & h & \fbox{$i$} \end{array} \right|. \label{col hom}
\end{align}

\section{Darboux transformations}

The process of transforming the pseudo-differential operator $L$ can be easily extended to the noncommutative case. This requires the following definition and theorem \cite{Oevel:Rogers}:

\begin{defn}
The function $\theta = \theta(x,t_q)$ is an eigenfunction for the hierarchy (\ref{commutater1}) if it satisfies the linear equations
\begin{align}
    \theta_{t_q} &= P_{\geq 1}(L^q)[\theta], \label{eigenfunctions}
\end{align}
which are compatible and can be considered simultaneously for each $q$.
\end{defn}
It is not assumed that $\theta$ and its $x$- and $t_q$-derivatives commute.
\begin{thm} \label{thm3}
Let $L$ satisfy \eqref{commutater1} and let $\psi= \psi(x,t_q)$ be a generic eigenfunction for this hierarchy. Then $\widetilde{L} = G_{\theta}LG_{\theta}^{-1}$ with
\begin{enumerate}
    \item[a)] $G_{\theta}[\psi] = \theta^{-1} \psi$, or
    \item[b)] $G_{\theta}[\psi] = (\theta_x)^{-1} \psi_x$, or a composition of the previous two transformations:
    \item[c)] $G_{\theta}[\psi] = \psi - \theta (\theta_x)^{-1} \psi_x$,
\end{enumerate}
satisfies the hierarchy $\widetilde{L}_{t_q} = [P_{\geq 1}(\widetilde{L}^q), \widetilde{L}]$
and $\tilde{\psi}_{t_q} = P_{\geq 1}(L^q)[\tilde{\psi}]$ where $\tilde{\psi} = G_{\theta}[\psi]$.
\end{thm}

It emerges that none of the three choices of $G_{\theta}$ transform the field $w$ in such a way that we can iterate the transformation and obtain quasideterminant solutions. We can, however, obtain quasideterminant structure for the function $f$, and the eigenfunction $\psi$, through the Darboux transformation $G_{\theta} = ((\theta^{-1})_x)^{-1}\partial_x \theta^{-1} = 1-\theta (\theta_x)^{-1} \partial_x$ given in Theorem \ref{thm3}c).

We note that quasideterminant structure is immediately evident from
\begin{align*}
G_{\theta}[\psi] &= \psi - \theta (\theta_x)^{-1} \psi_x = \left| \begin{array}{cc} \theta & \fbox{$\psi$} \\
\theta_x & \psi_x \end{array} \right|.
\end{align*}
Let $\theta_i$, $i=1, \ldots, n$ be a particular set of eigenfunctions and introduce the notation $\Theta = (\theta_1, \theta_2, \ldots, \theta_n)$.  To iterate the Darboux transformation, let $\theta_{[1]} = \theta_1$ and $\psi_{[1]} = \psi$ be a general eigenfunction of $L_{[1]}=L$.  Then $\psi_{[2]} := G_{\theta_{[1]}}[\psi_{[1]}]$ and $\theta_{[2]} = \psi_{[2]}|_{\psi \rightarrow \theta_2}$ are eigenfunctions for $L_{[2]} = G_{\theta_{[1]}} L_{[1]} G_{\theta_{[1]}}^{-1}$.  In general, for $n \geq 1$ we define the $n$th Darboux transformation of $\psi$ by

\begin{align*}
    \psi_{[n+1]} &= \psi_{[n]} - \theta_{[n]} (\theta_{[n]x})^{-1} \psi_{[n]x},
\end{align*}
in which
\begin{align*}
    \theta_{[k]} &= \psi_{[k]}|_{\psi \rightarrow \theta_k}.
\end{align*}

For example,

\begin{align*}
    \psi_{[2]} &= \psi - \theta_1 (\theta_{1x})^{-1} \psi_x =  \left| \begin{array}{cc} \theta_1 & \fbox{$\psi$} \\ \theta_1^{(1)} & \psi^{(1)} \end{array} \right|, \\
    \psi_{[3]} &= \psi - \theta_1 (\theta_{1x})^{-1} \psi_x -(\theta_2 - \theta_1 (\theta_{1x})^{-1} \theta_{2x})(\theta_2 - \theta_1 (\theta_{1x})^{-1} \theta_{2x})_x^{-1}(\psi - \theta_1 (\theta_{1x})^{-1} \psi_x)_x \\
    & = \left|  \begin{array}{ccc} \theta_1 & \theta_2 & \fbox{$\psi$} \\
    \theta_1^{(1)} & \theta_2^{(1)} & \psi^{(1)} \\ \theta_1^{(2)} & \theta_2^{(2)} & \psi^{(2)} \end{array} \right|,
\end{align*}
where ${}^{(k)}$ denotes the $k$th $x$-derivative. After $n$ iterations we have

\begin{align*}
    \psi_{[n+1]} &= \left| \begin{array}{cc} \Theta & \fbox{$\psi$} \\
    \vdots & \vdots \\
    \Theta^{(n-1)} & \psi^{(n-1)} \\
    \Theta^{(n)} & \psi^{(n)}
    \end{array} \right|.
\end{align*}
Next we determine the Darboux-transformed fields $\tilde{w}, \tilde{w}_s (s=1,2, \ldots)$ by calculating
\begin{align*}
    \widetilde{L} &= ((\theta^{-1})_x)^{-1} \partial_x \theta^{-1} L
    \theta \partial_x^{-1} (\theta^{-1})_x \\
    &= \partial_x - (-\theta (\theta_x)^{-1} f)_x(- \theta (\theta_x)^{-1} f)^{-1}
    + \left( \frac{1}{2}(-\theta (\theta_x)^{-1} f)_{xx} (-\theta (\theta_x)^{-1} f)^{-1}
    \right. \\& \left. \quad -(-\theta (\theta_x)^{-1} f)_x(-\theta (\theta_x)^{-1} f)^{-1} (-\theta (\theta_x)^{-1} f)_x(-\theta (\theta_x)^{-1} f)^{-1}
    - \frac{1}{2} ((-\theta (\theta_x)^{-1} f)^{-1})_y (-\theta (\theta_x)^{-1} f)^{-1} \right) \partial_x^{-1} + \ldots,
\end{align*}
which leaves the ncmKP hierarchy invariant, preserving the structure of $w$, $w_s, (s=1,2, \ldots)$.  The
coefficients
\begin{align*}
    \tilde{w} &=  - (-\theta (\theta_x)^{-1} f)_x(- \theta (\theta_x)^{-1} f)^{-1}, \\
    \tilde{w}_1 &= \frac{1}{2}(-\theta (\theta_x)^{-1} f)_{xx} (-\theta (\theta_x)^{-1} f)^{-1}
    -(-\theta (\theta_x)^{-1} f)_x(-\theta (\theta_x)^{-1} f)^{-1} (-\theta (\theta_x)^{-1} f)_x(-\theta (\theta_x)^{-1} f)^{-1}
    \\ & \quad - \frac{1}{2} ((-\theta (\theta_x)^{-1} f)^{-1})_y (-\theta (\theta_x)^{-1} f)^{-1},  \\
    {} & \cdots {} \nonumber
\end{align*}
will satisfy (\ref{hierarchy8}) and (\ref{hierarchy9}).  In particular, $\tilde{w}$ will satisfy the ncmKP
equation.  Using the fact that $\tilde{w}$ is of the form $- \tilde{f}_x \tilde{f}^{-1}$, we obtain

\begin{align*}
    \tilde{f} &= - \theta (\theta_x)^{-1} f = \left| \begin{array}{cc} \theta & \fbox{0} \\
    \theta_x & 1 \end{array} \right| f.
\end{align*}
If we let $f = f_{[1]}$, then for the $n$th Darboux transformation of $f$ we have
\begin{align*}
    f_{[n+1]} &= \left| \begin{array}{cc} \Theta & \fbox{$0$} \\
    \vdots & \vdots \\
    \Theta^{(n-1)} & 0 \\
    \Theta^{(n)} & 1
    \end{array} \right|f.
\end{align*}
We note that an analogous transformation can be made by letting $g = f^{-1}$, so that

\begin{align*}
w = -(g^{-1})_x g = g^{-1} g_x
\end{align*}
satisfies the ncmKP equation.  For the function $g$ we get

\begin{align*}
    \tilde{w} &= (g \theta_x \theta^{-1})^{-1} (g \theta_x \theta^{-1})_x,
\end{align*}
so that

\begin{align*}
    \tilde{g} &= g \theta_x \theta^{-1} = - g \left| \begin{array}{cc}  \theta & 1 \\
    \theta_x & \fbox{$0$} \end{array} \right|.
\end{align*}
If we let $g = g_{[1]}$, then for the $n$th Darboux transformation of $g$ we have
\begin{align*}
    g_{[n+1]} &= - g \begin{vmatrix}
    \Theta&1\\
    \Theta^{(1)}&0\\
    \vdots&\vdots\\
    \Theta^{(n-1)}&0\\
    \Theta^{(n)}&\fbox{$0$}
    \end{vmatrix}.
\end{align*}

\section{Derivatives of quasiwronskians}

Derivatives of quasideterminants have been considered in\cite{claire:jon, MR2321659}.  Let $\widehat{\Theta} = \left(
\theta_j^{(i-1)} \right)_{i,j=1,\ldots,n}$ be the $n \times n$ wronskian matrix of $\theta_1, \ldots ,
\theta_n$, where $^{(k)}$ denotes the $k$th derivative and let $e_k$ be the $n$-vector $(\delta_{ik})$ (i.e. a
column vector with $1$ in the $k$th row and $0$ elsewhere).  We consider derivatives of the form

\begin{align}
    Q(i,j) &= \left| \begin{array}{cc} \widehat{\Theta} & e_{n-j} \\ \Theta^{(n+i)} & \fbox{$0$} \end{array} \right|.
\end{align}
Assuming $n$ is arbitrarily large, we may summarise the properties of $Q(i,j)$ as (see \cite{claire:jon})
\begin{align}
    Q(i,j) &= \left\{ \begin{array}{ll} -1 & i+j+1 = 0 \\ 0 & (i < 0 \: \textrm{or} \: j < 0) \: \textrm{and} \:
    i+j+1 \neq 0 \end{array} \right. . \label{Q con}
\end{align}
We call this type of quasideterminant a \textit{quasiwronskian}.  If we relabel and rescale the variables so that $x_1 = x$, $x_2 = y$, $x_3 = -4t$, $\Theta$ satisfies the linear equations

\begin{align}
    \Theta_{x_2} &= \Theta_{xx}  \nonumber , \\
    \Theta_{x_3} &= \Theta_{xxx}.
\end{align}
We may allow $\Theta$ to depend on higher variables $x_k$ and impose the natural dependence $\Theta_{x_k} =
\Theta_{\underbrace{x \cdots x}_k}$.

Using the conditions (\ref{Q con}) we obtain (see \cite{claire:jon})

\begin{align}
    \frac{\partial}{\partial x_m} Q(i,j) &= Q(i+m,j) - Q(i,j+m) + \sum_{k=0}^{(m-1)} Q(i,k) Q(m-k-1,j). \label{Q der2}
\end{align}
This formula is known in the commutative case \cite{MR1041529, MR1776504} but arises in connection with the construction of the KP hierarchy rather than its solutions.

In addition to $Q(i,j)$ we can define a shifted version, which we will call $Q'(i,j)$
\begin{align*}
    Q'(i,j) &= \left| \begin{array}{cc} \Theta^{(1)} & 0 \\ \vdots & \vdots \\ \Theta^{(n-j)} & 1 \\ \vdots & \vdots \\
    \Theta^{(n)} & 0 \\ \Theta^{(n+i+1)} & \fbox{$0$} \end{array} \right| .
\end{align*}
This satisfies an equation similar to (\ref{Q der2}).

\section{Direct verification}
In this section we derive identities which link the quasideterminant solutions of ncKP and ncmKP.  Note that the Lax pairs of ncKP and ncmKP are the same when the vacuum solutions are trivial. Let $\Theta$ be a common
eigenfunction for these two (trivial vacuum) Lax pairs. We find that the solutions
of ncmKP are

\[
w= -F_x F^{-1}, \qquad W=-F_y F^{-1}
\]
or equivalently
\[
w=  G^{-1} G_x, \qquad W= G^{-1} G_y,
\]
where
\begin{equation}\label{F}
 F=\begin{vmatrix}
    \Theta&\fbox{$0$}\\
    \Theta^{(1)}&0\\
    \vdots&\vdots\\
    \Theta^{(n-1)}&0\\
    \Theta^{(n)}&1
    \end{vmatrix}, \quad
    G=F^{-1}=\begin{vmatrix}
    \Theta&1\\
    \Theta^{(1)}&0\\
    \vdots&\vdots\\
    \Theta^{(n-1)}&0\\
    \Theta^{(n)}&\fbox{$0$}
    \end{vmatrix}.
\end{equation}
The inverse of $F$ is obtained from the expression for $F$ by swapping the boxed entry and the 1 in the last column of $F$.

In our discussion we also use the solutions $v=-2 Q$ and $\hat v=-2 Q'$ of ncKP equation \cite{claire:jon}
\begin{equation}\label{kpp}
    (v_t+v_{xxx}+3 v_x v_x)_x+3 v_{yy}-3 [v_x,v_y]=0.
\end{equation}
Here, for convenience this is written in potential form (the usual KP variable is $u=v_x$).
\[
    Q=Q(0,0)=\begin{vmatrix}
    \Theta&0\\
    \Theta^{(1)}&0\\
    \vdots&\vdots\\
    \Theta^{(n-1)}&1\\
    \Theta^{(n)}&\fbox{$0$}
    \end{vmatrix},\quad
    Q'=Q'(0,0)=\begin{vmatrix}
    \Theta^{(1)}&0\\
    \Theta^{(2)}&0\\
    \vdots&\vdots\\
    \Theta^{(n)}&1\\
    \Theta^{(n+1)}&\fbox{$0$}
    \end{vmatrix}.
\]
Note that $Q'$ is only a solution if the vacuum is trivial.

In a similar way we define
\begin{align}
    F(j)=\begin{vmatrix}
    \Theta&\fbox{$0$}\\
    \Theta^{(1)}&0\\
    \vdots&\vdots\\
    \Theta^{(n-j)}&1\\
    \vdots&\vdots\\
    \Theta^{(n)}&0
    \end{vmatrix}, \label{row diff} \quad
    G(j)=\begin{vmatrix}
    \Theta&1\\
    \Theta^{(1)}&0\\
    \vdots&\vdots\\
    \Theta^{(n-1)}&0\\
    \Theta^{(n+j)}&\fbox{$0$}
    \end{vmatrix},
\end{align}
and note that $F=F(0)$ and $G=G(0)$. Using \eqref{col hom}, we have homological relations expressed as the identities
\begin{equation}\label{FQ id}
    FQ(0,j)=F(j+1)
\end{equation}
and
\begin{equation}\label{FQhat id}
    Q'(j,0)G=-G(j+1).
\end{equation}

Now consider the derivatives of $F(j)$: using \eqref{row diff} and \eqref{FQ id},
\begin{equation}
    F(j)_x=FQ'(0,j)-F(j+1)=F(Q'(0,j)-Q(0,j)).
\end{equation}
More generally, if we assume that $\Theta$ satisfies the linear PDEs
$\Theta_{x_k}=\Theta_{\underbrace{x\cdots x}_k}$, we have
\begin{align}
    F(j)_{x_{k+1}}&=\sum_{i=0}^{k}F(i)Q'(k-i,j)-F(k+j+1)\\
    &=F\left(Q'(k,j)+\sum_{i=1}^{k} Q(0,i-1)Q'(k-i,j)-Q(0,k+j)\right).
\end{align}
Thus
\begin{align*}
    F_{x}&=FQ'-F(1)=F(Q'-Q),\\
    F_{xx}
    &=F\bigl((Q'-Q)^2+Q'_x-Q_x\bigr),
\end{align*}
and
\begin{equation*}
    F_y=FQ'(1,0)+F(1)Q'-F(2),
\end{equation*}
and so
\begin{equation}
    F_{xx}+F_y=2FQ'_x.
\end{equation}
Using a Jacobi identity and \eqref{FQhat id} we can show that;
\begin{align*}
  Q'(0,1)&=\begin{vmatrix}
    \Theta^{(1)}&0\\
    \vdots&\vdots\\
    \Theta^{(n-1)}&1\\
    \Theta^{(n)}&0\\
    \Theta^{(n+1)}&\fbox{$0$}
    \end{vmatrix}=\begin{vmatrix}
    \Theta&1&0\\
    \Theta^{(1)}&0&0\\
    \vdots&\vdots&\vdots\\
    \Theta^{(n-1)}&0&1\\
    \Theta^{(n)}&0&0\\
    \Theta^{(n+1)}&0&\fbox{$0$}
    \end{vmatrix}=Q(1,0)-G(1)F^{-1}Q=Q(1,0)+Q'Q.
\end{align*}
This is the noncommutative version of the first bilinear identity in the ncmKP hierarchy. This  can also be obtained from expanding $F_{xx}$ in two different ways. This noncommutative  identity can be generalized to get to the other members of the hierarchy;
\begin{equation}\label{mKP hierarchy}
  Q'(i,j)=Q(i+1,j-1)+Q'(i,0)Q(0,j-1).
\end{equation}
This follows immediately from considering $Q'(i,j)$ written as
\begin{align*}
\begin{vmatrix}
    \Theta&1&0\\
    \Theta^{(1)}&0&0\\
    \vdots&\vdots&\vdots\\
    \Theta^{(n-j)}&0&1\\
    \vdots&\vdots&\vdots\\
    \Theta^{(n-1)}&0&0\\
    \Theta^{(n)}&0&0\\
    \Theta^{(n+1+i)}&0&\fbox{$0$}
    \end{vmatrix}&=\begin{vmatrix}
    \Theta&0\\
    \Theta^{(1)}&0\\
    \vdots&\vdots\\
    \Theta^{(n-j)}&1\\
    \vdots&\vdots\\
    \Theta^{(n-1)}&0\\
    \Theta^{(n+1+i)}&\fbox{$0$}
    \end{vmatrix}-\begin{vmatrix}
    \Theta&1\\
    \Theta^{(1)}&0\\
    \vdots&\vdots\\
    \Theta^{(n-j)}&0\\
    \vdots&\vdots\\
    \Theta^{(n-1)}&0\\
    \Theta^{(n+1+i)}&\fbox{$0$}
    \end{vmatrix}\begin{vmatrix}
    \Theta&1\\
    \Theta^{(1)}&0\\
    \vdots&\vdots\\
    \Theta^{(n-j)}&0\\
    \vdots&\vdots\\
    \Theta^{(n-1)}&0\\
    \Theta^{(n)}&\fbox{$0$}
    \end{vmatrix}^{-1}\begin{vmatrix}
    \Theta&0\\
    \Theta^{(1)}&0\\
    \vdots&\vdots\\
    \Theta^{(n-j)}&1\\
    \vdots&\vdots\\
    \Theta^{(n-1)}&0\\
    \Theta^{(n)}&\fbox{$0$}
    \end{vmatrix}\\
    &=Q(i+1,j-1)-G(i+1)F^{-1}Q(0,j-1)
\end{align*}
and then using \eqref{FQhat id}.

In the commutative case, \eqref{mKP hierarchy} becomes (in Frobenius notation)
\[
\hat \tau_{(i|j)}\tau-\hat\tau_{(i|0)}\tau_{(0|j-1)}+\hat\tau\tau_{(i+1|j-1)}=0.
\]

We substitute $w, W$ and their derivatives into equations (\ref{NCmKPequation}) and (\ref{ncmKpextra}).  Equation (\ref{ncmKpextra}) is satisfied straightforwardly and equation (\ref{NCmKPequation}) is satisfied only after application of the identities (\ref{FQ id}, \ref{FQhat id}, \ref{mKP hierarchy}).

A Miura transformation \cite{Dimakis:Muller-Hoissen2006}, taking us from a solution of the ncmKP to that of the ncKP can be obtained from the Gelfand-Dikii approach.  The transformation takes the form:
\begin{align}\label{MT1}
     -w_x- w^2+W= F u F^{-1}, \qquad u&=v_x,
\end{align}
where  $v$ is a solution of the ncKP equation in potential form (\ref{kpp}) and $w,W,F$ are the fields from the
ncmKP. Notice here, that in the commuting case the fields $F$ and $F^{-1}$ cancel to give
\begin{align*}
     -w_x- w^2+W= u , \qquad u&=v_x,\qquad  W_x=w_y.
\end{align*}
This is the usual Miura transformation.  For the noncommuting case there is a second Miura transformation
\begin{align}\label{MT2}
    w_x- w^2+W= F \hat{u} F^{-1}, \qquad \hat{u}&=\hat{v}_x,
\end{align}
which relates the ncmKP with a different solution $\hat{v}$ of the ncKP. Both (\ref{MT1}) and (\ref{MT2}) can be
directly verified by using the quasiwronskian form of the functions as given earlier in the paper.

\section{Conclusions}
In this paper we have investigated a Gelfand-Dikki approach to a noncommutative modified KP hierarchy. The
equations obtained are similar to those obtained in the commuting version but have additional terms that are
commutators.  A construction of the noncommutative mKP equation was given and it was seen to match up with that of Wang and Wadati \cite{Wang:Wadati}.  We have shown that quasiwronskian solutions can be built up by means of Darboux
transformations. Additionally we have used direct methods to show that these solutions satisfy the ncmKP
equations using quasideterminantal identities.  As with the work on the ncKP equation \cite{claire:jon} we have
not at any point specified what kind of noncommutative objects we are looking at.  This means that our results
will hold for any noncommutative situation. For instance we could be looking at a matrix version of mKP or a
quaternionic version.

\bibliographystyle{plain}
\bibliography{final_ncmkp}

\end{document}